\documentclass[doublecol,linenumbers]{epl2} 

\title{How to suppress the backscattering of conduction electrons?}
\shorttitle{How to suppress the backscattering of conduction electrons?} 

\author{O. V. Kibis}
\shortauthor{O. V. Kibis}

\institute{Department of Applied and Theoretical Physics,
Novosibirsk State Technical University, Karl Marx Avenue 20,
630073 Novosibirsk, Russia}


\abstract{It is shown theoretically that the strong coupling of
electrons to a high-frequency electromagnetic field results in the
nulling of electron backscattering within the Born approximation.
The conditions of the effect depend only on field parameters and
do not depend on concrete form of scattering potential. As a
consequence, this phenomenon is of universal physical nature and
can take place in various conducting systems. Since the
suppression of electron backscattering results in decreasing
electrical resistance, the solved quantum-mechanical problem opens
a new way to control electronic transport properties of conductors
by a laser-generated field. Particularly, the elaborated theory is
applicable to nanostructures exposed to a strong monochromatic
electromagnetic wave.}

\begin{document}

\maketitle

\section{Introduction}
The backscattering is fundamental physical process consisting in
the reflection of moving electrons back to the direction from
which they came. Resulting in dissipation of kinetic momentum of
the electrons, the backscattering lies in the core of all
mechanisms of electrical resistance in conductors. Therefore, the
seeking for ways to suppress the backscattering is problem of both
general physical interest and rich applied capabilities. In the
given Letter, it is shown theoretically how the backscattering of
conduction electrons can be suppressed by a high-frequency
electromagnetic field.

Due to advances in laser physics, a strong electromagnetic field
became an ordinary tool to manipulate electronic properties of
various systems. In contrast to the case of weak electromagnetic
field, the interaction between electrons and the strong field
cannot be described as a perturbation. Therefore, the system
``electron + strong electromagnetic field'' should be considered
as a bound electron-field object which was called ``electron
dressed by field'' (dressed electron)
\cite{Cohen-Tannoudji_b98,Scully_b01}. In atomic and molecular
systems, the interaction between electrons and a dressing field
results in various phenomena, including both the photon-induced
modification of electron energy spectrum (the dynamic Stark
effect) \cite{Autler_55,Chini_12,Yu_13,Demekhin_13,Giner_13} and
the photon-assisted scattering of electrons by atoms and molecules
\cite{Bunkin_66,Kroll_73,Kanya_10,Bhatia_12,Flegel_13}. In
condensed matters, research activity was focused on the dynamic
Stark effect
\cite{Elesin_69,Mysyrowicz_86,Koch_88,Vu_04,Faist_05,Lopez_08,Kibis_09,Kibis_10,Kibis_11_1,Savelev_11,Kibis_11,Kibis_12,Hayat_12,Kibis_13}
and the photon-assisted non-stationary quantum transport through
potential barriers \cite{Pedersen_98,
Moskalets_02,Platero_04,Moskalets_book}. As to effect of dressing
field on scattering processes which are responsible for stationary
(dc) electronic transport in usual conducting systems, it escaped
attention before. Filling this gap in the theory, I unexpectedly
found that the modification of electron wave functions by a strong
high-frequency electromagnetic field results in the nulling of
scattering matrix elements. As a consequence, the phenomenon
announced above appears. Surprisingly, this quantum effect can be
derived directly from the basic principles of quantum mechanics
with using simple pen-and-paper calculations as follows.

\section{The model} For definiteness, we will restrict our
consideration by the case of alternating homogeneous electric
field but the proper generalization for any electromagnetic field
can be easily made. Let an electron be subjected to such an
electric field
\begin{equation}\label{E}
\mathbf{E}(t)=\mathbf{E}_0\sin\omega t,
\end{equation}
where $E_0$ is the amplitude of the field, and $\omega$ is the
frequency of the field. In the absence of scatterers, the wave
function of the electron, $\psi$, satisfies the Schr\"odinger
equation
\begin{equation}\label{shr}
i\hbar\frac{\partial\psi}{\partial t}=\hat{\cal{H}}_0\psi
\end{equation}
with the Hamiltonian
\begin{equation}\label{H0}
\hat{\cal{H}}_0=\frac{1}{2m}\left[\hat{\mathbf{p}}-\frac{e}{c}\mathbf{A}(t)\right]^2,
\end{equation}
where $\hat{\mathbf{p}}$ is the operator of electron momentum, $m$
is the effective electron mass in a conductor, $e$ is the electron
charge,
\begin{equation}\label{A}
\mathbf{A}(t)=\frac{c\mathbf{E}_0}{\omega}\cos\omega t
\end{equation}
is the vector potential of the field, and the field frequency
$\omega$ is assumed to be far from resonant electron frequencies
corresponding to interband electron transitions in the conductor.
It should be noted that the Hamiltonian (\ref{H0}) with the same
vector potential (\ref{A}) describes a two-dimensional (or
one-dimensional) electron system subjected to a plane linearly
polarized monochromatic electromagnetic wave with the frequency
$\omega$ and the amplitude $E_0$, which propagates perpendicularly
to the system. As a consequence, the theory developed below is
applicable, particularly, to such nanostructures as quantum wells
and quantum wires exposed to the electromagnetic wave.

Since the vector potential (\ref{A}) does not depend on
coordinates, the nonstationary Schr\"odinger equation (\ref{shr})
with the Hamiltonian (\ref{H0}) can be solved accurately. Namely,
let us seek the wave function $\psi$ in the form
\begin{equation}\label{p}
\psi=V^{-1/2}\exp\left[-iF(t)-i\mathbf{k}\mathbf{r}\right],
\end{equation}
where $\mathbf{k}$ is the electron wave vector, $\mathbf{r}$ is
the electron radius-vector, $V$ is the normalization volume, and
$F(t)$ is the required function. Substituting the wave function
(\ref{p}) into the Schr\"odinger equation (\ref{shr}), we arrive
at the differential equation,
$$
\hbar\frac{dF(t)}{dt}=\frac{\hbar^2\mathbf{k}^2}{2m}-\frac{\hbar
e}{mc} \mathbf{k}\mathbf{A}(t)+\frac{e^2}{2mc^2}\mathbf{A}^2(t),
$$
which can be easily solved by direct integration over time $t$. As
a result, the exact wave function of the dressed electron
(\ref{p}) can be written as
\begin{eqnarray}\label{psi}
\psi_\mathbf{k}(\mathbf{r},t)&=&\exp\left[-i\left(\frac{\varepsilon_k
t}{\hbar}+\frac{E_0^2e^2t}{4m\omega^2\hbar}+\frac{E_0^2e^2}{8m\omega^3\hbar}\sin2\omega
t\right.\right.\nonumber\\
&-&\left.\left.\frac{e\mathbf{E}_0\mathbf{k}}{m\omega^2}\sin\omega
t\right)\right]\varphi_{\mathbf{k}}(\mathbf{r}),
\end{eqnarray}
where
$\varphi_{\mathbf{k}}(\mathbf{r})={V}^{-1/2}\exp(i\mathbf{k}\mathbf{r})$
is the plane electron wave, and $\varepsilon_k={\hbar^2k^2}/{2m}$
is the energy spectrum of bare electron. As expected, in the
absence of the field ($E_0=0$) the wave function of dressed
electron (\ref{psi}) turns into the wave function of bare
electron,
\begin{equation}\label{psi0}
\psi_{\mathbf{k}}^{(0)}(\mathbf{r},t)=\varphi_{\mathbf{k}}(\mathbf{r})\exp(-i\varepsilon_kt/\hbar).
\end{equation}
It should be stressed that the velocity of the dressed electron
averaged over the field period $T=2\pi/\omega$,
$$\mathbf{v}(\mathbf{k})=\frac{1}{T}\int_0^T\left\langle\psi_\mathbf{k}(\mathbf{r},t)\left|\frac{\hat{\mathbf{p}}-e\mathbf{A}(t)/c}{m}
\right|\psi_\mathbf{k}(\mathbf{r},t)\right\rangle dt=\frac{\hbar
\mathbf{k}}{m},$$ exactly coincides with the stationary velocity
of bare electron in the state (\ref{psi0}) with the same wave
vector $\mathbf{k}$,
$$\mathbf{v}^{(0)}(\mathbf{k})=\left\langle\psi_{\mathbf{k}}^{(0)}(\mathbf{r},t)\left|\frac{\hat{\mathbf{p}}}{m}
\right|\psi_{\mathbf{k}}^{(0)}(\mathbf{r},t)\right\rangle=\frac{\hbar
\mathbf{k}}{m}.$$ Thus, the high-frequency field does not change
the mean electron velocity and, correspondingly, does not
influence directly on stationary (dc) electronic transport.
However, the formal mathematical difference between the wave
function of dressed electron (\ref{psi}) and the wave function of
bare electron (\ref{psi0}) effects on scattering processes
discussed below.

Let an electron moves in a conductor with scatterers in the
presence of the same field (\ref{A}). Then the wave function of
the conduction electron, $\Psi(\mathbf{r},t)$, satisfies the
Schr\"odinger equation
\begin{equation}\label{SE}
i\hbar\frac{\partial\Psi(\mathbf{r},t)}{\partial t}=[\hat{\cal
H}_0+U(\mathbf{r})]\Psi(\mathbf{r},t),
\end{equation}
where the total scattering potential $U(\mathbf{r})$ is
superposition of potentials arisen from macroscopically large
number of scatterers in the conductor. In what follows, we will
assume the scattering potential energy $U(\mathbf{r})$ to be a
small perturbation. This allows to apply the conventional
perturbation theory \cite{Landau_3} to describe the electron
scattering by the potential $U(\mathbf{r})$. Since the functions
(\ref{psi}) with different wave vectors $\mathbf{k}$ form the
complete function system for any time $t$, we can seek solutions
of the Schr\"odinger equation (\ref{SE}) as an expansion
\begin{equation}\label{P}
\Psi(\mathbf{r},t)=\sum_{\mathbf{k}^\prime}a_{\mathbf{k}^\prime}(t)\psi_{\mathbf{k}^\prime}(\mathbf{r},t).
\end{equation}
It should be stressed that eq.~(\ref{psi}) gives exact wave
functions of dressed electron. Therefore, the using of the basis
(\ref{psi}) in the expansion (\ref{P}) takes into account the
interaction between an electron and the dressing field (\ref{A})
in full, i.e. non-perturbatively. Let an electron be in the state
(\ref{psi}) with the wave vector $\mathbf{k}$ at the time $t=0$.
Correspondingly,
$a_{\mathbf{k}^\prime}(0)=\delta_{\mathbf{k}^\prime,\mathbf{k}}$,
where $\delta_{\mathbf{k}^\prime,\mathbf{k}}$ is the Kronecker
symbol. Substituting the expansion (\ref{P}) into the
Schr\"odinger equation (\ref{SE}) and restricting the accuracy by
the first order of the perturbation theory, we arrive at the
expression
\begin{equation}\label{ak}
a_{\mathbf{k}^\prime}(t)=-i\frac{U_{\mathbf{k}^\prime\mathbf{k}}}{\hbar}
\int_0^te^{i(\varepsilon_{k^\prime}-\varepsilon_k)t^\prime/{\hbar}}e^{if_{\mathbf{k}^\prime\mathbf{k}}\sin\omega
t^\prime}dt^\prime,
\end{equation}
where $\mathbf{k}^\prime\neq\mathbf{k}$,
\begin{equation}\label{U}
U_{\mathbf{k}^\prime\mathbf{k}}=\left\langle\varphi_{\mathbf{k}^\prime}(\mathbf{r})
\left|U(\mathbf{r})\right|\varphi_{\mathbf{k}}(\mathbf{r})\right\rangle
\end{equation}
is the matrix element of the scattering potential, and
\begin{equation}\label{f}
f_{\mathbf{k}^\prime\mathbf{k}}=\frac{e\mathbf{E}_0
(\mathbf{k}-\mathbf{k}^\prime)}{m\omega^2}.
\end{equation}
To proceed analysis of the problem, we have to slightly extend the
conventional perturbation theory \cite{Landau_3} which was
developed by Dirac for stationary basis wave functions, taking
into account the non-stationarity of the wave functions
(\ref{psi}). Namely, let us apply the Jacobi-Anger expansion
\cite{Gradstein},
$$e^{iz\sin\theta}=\sum_{n=-\infty}^{\infty}J_n(z)e^{in\theta},$$
in order to rewrite eq.~(\ref{ak}) as
\begin{eqnarray}\label{wk}
|a_{\mathbf{k}^\prime}(t)|^2&=&\frac{\left|U_{\mathbf{k}^\prime\mathbf{k}}\right|^2}{\hbar^2}
\Bigg|\sum_{n=-\infty}^{\infty}{J_n\left(f_{\mathbf{k}^\prime\mathbf{k}}\right)}\,
e^{i(\varepsilon_{k^\prime}-\varepsilon_k+n\hbar\omega)t/{2\hbar}}\nonumber\\
&\times&\int_{-t/2}^{t/2}e^{i(\varepsilon_{k^\prime}-\varepsilon_k
+n\hbar\omega)t^\prime/\hbar}dt^\prime\Bigg|^2,
\end{eqnarray}
where $J_n(z)$ is the Bessel function of the first kind. Since the
integrals in eq.~(\ref{wk}) for long time $t$ turn into the
delta-function
$$\delta(\varepsilon)=
\frac{1}{2\pi\hbar}\lim_{t\rightarrow\infty}
\int_{-t/2}^{t/2}e^{i\varepsilon t^\prime/\hbar}dt^\prime,$$ the
expression (\ref{wk}) takes the form
\begin{equation}\label{wkk}
|a_{\mathbf{k}^\prime}(t)|^2=4\pi^2\left|U_{\mathbf{k}^\prime\mathbf{k}}\right|^2
\sum_{n=-\infty}^{\infty}J_n^2\left(f_{\mathbf{k}^\prime\mathbf{k}}\right)
\delta^2(\varepsilon_{k^\prime}-\varepsilon_k+n\hbar\omega).
\end{equation}
To transform square delta-functions in eq.~(\ref{wkk}), we can
apply the conventional procedure (see, e.g.,
ref.~\cite{Landau_4}),
$$\delta^2(\varepsilon)=\delta(\varepsilon)\delta(0)
=\frac{\delta(\varepsilon)}{2\pi\hbar}\lim_{t\rightarrow\infty}
\int_{-t/2}^{t/2}e^{i0\times
t^\prime/\hbar}dt^\prime=\frac{\delta(\varepsilon)t}{2\pi\hbar}.$$
Then the probability of the electron scattering between the states
(\ref{psi}) with the wave vectors $\mathbf{k}$ and
$\mathbf{k}^\prime$ per unit time,
$w_{\mathbf{k}^\prime\mathbf{k}}=d|a_{\mathbf{k}^\prime}(t)|^2/d
t$, is given by
\begin{equation}\label{W}
w_{\mathbf{k}^\prime\mathbf{k}}=\frac{2\pi}{\hbar}
\left|U_{\mathbf{k}^\prime\mathbf{k}}\right|^2\sum_{n=-\infty}^{\infty}J_n^2\left(f_{\mathbf{k}^\prime\mathbf{k}}\right)
\delta(\varepsilon_{k^\prime}-\varepsilon_k+n\hbar\omega).
\end{equation}

\section{Results} The announced effect follows from the
fact that all terms in the probability (\ref{W}) can be nulled
with a strong high-frequency field. First of all, let us consider
the terms with $n\neq0$. Physically, they describe the processes
of collisional (intraband) absorption and emission of photons by
conduction electron. Within the classical Drude theory, the
collisional absorption of the field (\ref{E}) by conduction
electrons is given by the well-known expression
\begin{equation}\label{Drude}
Q=\frac{1}{T}\int_0^T
\mathbf{j}(t)\mathbf{E}(t)dt=\frac{E_0^2}{2}\frac{\sigma_0}{1+(\omega\tau)^2},
\end{equation}
where $Q$ is the period-averaged field energy absorbed by
conduction electrons per unit time and per unit volume,
$\mathbf{j}(t)$ is the ohmic current density induced by the
high-frequency field $\mathbf{E}(t)$, $\sigma_0=e^2n\tau/m$ is the
static Drude conductivity, $n$ is the density of conduction
electrons, and $\tau$ is the electron relaxation time (see, e.g.,
Refs.~\cite{Ashcroft,Harrison}). Evidently, the Drude optical
absorption (\ref{Drude}) is negligibly small under the condition
\begin{equation}\label{tau}
\omega\tau\gg1.
\end{equation}
Therefore, it it reasonable to expect that terms with $n\neq0$ in
eq.~(\ref{W}), which describe the same optical absorption from the
quantum point of view, are also negligibly small under the
condition (\ref{tau}).

Within quantum theory, the absence of energy exchange between
conduction electrons and a high-frequency electromagnetic field
follows directly from the conservation laws. Namely, it is
well-known that the conservation laws for energy and momentum
forbid the processes of intraband absorption and emission of
photons by conduction electrons. These processes are appreciable
only if the scattering-induced washing of electron energy,
$\delta\varepsilon$, is comparable with the photon energy
$\hbar\omega$. As a consequence, the terms with $n\neq0$ in
eq.~(\ref{W}) can be neglected if the field frequency $\omega$ is
high to satisfy the condition $\hbar\omega\gg\delta\varepsilon$.
Applying the uncertainty relation for energy,
$\delta\varepsilon\sim\hbar/\tau$, to this condition, we arrive
again at the inequality (\ref{tau}).

The simple arguments based on the conservation laws can be
verified formally by analysis of the Schr\"odinger equation.
Namely, we have to consider the Hamiltonian $\hat{\cal
H}=\hat{\mathbf{p}}/2m+U(\mathbf{r})$ describing a conduction
electron in a scattering potential $U(\mathbf{r})$ in the absence
of the field. The stationary Schr\"odinger equation with the
Hamiltonian, $\hat{\cal
H}u_l(\mathbf{r})=\varepsilon_lu_l(\mathbf{r})$, yields the energy
spectrum $\varepsilon_l$ and the wave functions $u_l(\mathbf{r})$
of the electron, where the index $l$ labels various eigenstates of
the Hamiltonian. Since the stationary eigenfunctions
$u_l(\mathbf{r})$ form the complete function system, the wave
function of free conduction electron,
$\varphi_{\mathbf{k}}(\mathbf{r})=V^{-1/2}\exp(i\mathbf{k}\mathbf{r})$,
can be expanded on the eigenfunctions as
$\varphi_{\mathbf{k}}(\mathbf{r})=\sum_lb_{\mathbf{k}l}u_l(\mathbf{r})$.
Then the matrix element (\ref{U}) is
\begin{equation}\label{M}
U_{\mathbf{k}^\prime\mathbf{k}}=\left\langle\varphi_{\mathbf{k}^\prime}(\mathbf{r})
\left|\hat{\cal
H}\right|\varphi_{\mathbf{k}}(\mathbf{r})\right\rangle
=\sum_l\varepsilon_lb^\ast_{\mathbf{k}^\prime l}b_{\mathbf{k}l}.
\end{equation}
Physically, the expansion coefficients $b_{\mathbf{k}l}$ describe
the broadening of energy spectrum of conduction electron caused by
the scattering: An electron with the wave vector $\mathbf{k}$ has
the non-zero probability $W_{\mathbf{k}l}=|b_{\mathbf{k}l}|^2$ to
be in the eigenstate $l$ with the energy $\varepsilon_l$. If to
rewrite the coefficients as
$b_{\mathbf{k}l}=\sqrt{W_{\mathbf{k}^\prime l}}e^{i\phi_{kl}}$,
the matrix element (\ref{M}) takes the form
\begin{equation}\label{in}
U_{\mathbf{k}^\prime\mathbf{k}}=\sum_l\varepsilon_l\sqrt{
W_{\mathbf{k}^\prime
l}W_{\mathbf{k}l}}e^{i(\phi_{kl}-\phi_{k^\prime l})}.
\end{equation}
\begin{figure}[th]
\includegraphics[width=0.48\textwidth]{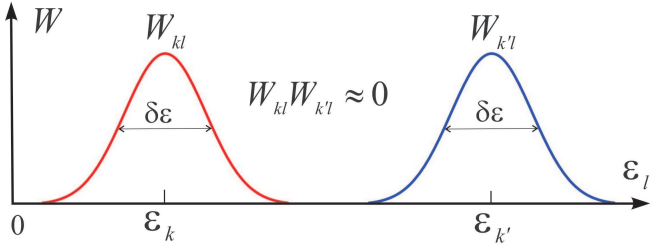}
\caption{Scheme of the scattering-induced washing of electron
states $\varepsilon_k$ and $\varepsilon_{k^\prime}$ in a
conductor. The overlap of the plotted probabilities $W_{kl}$ and
$W_{k^{\prime}l}$
--- and, correspondingly, the product $W_{kl}W_{k^{\prime}l}$ ---
is negligibly small under the condition
$|\varepsilon_k-\varepsilon_{k^\prime}|\gg\delta\varepsilon$.}
\end{figure}
The scattering-induced washing of electron energy,
$\delta\varepsilon$, is the energy range near the energy of free
electron, $\varepsilon_k={\hbar^2k^2}/{2m}$, where the probability
$W_{\mathbf{k}l}$ distinctly differs from zero (see fig.~1).
Therefore, the product $W_{\mathbf{k}^\prime l}W_{\mathbf{k}l}$
--- and, correspondingly, the sum in eq.~(\ref{in})
--- is very small under the condition
$|\varepsilon_k-\varepsilon_{k^\prime}|\gg\delta\varepsilon$. As a
consequence, the matrix element (\ref{in}) is negligibly small in
those terms of eq.~(\ref{W}), which satisfy this condition. Since
the terms with $n\neq0$ in eq.~(\ref{W}) describe the electron
scattering between the states $\mathbf{k}$ and $\mathbf{k}^\prime$
with the energy difference
$|\varepsilon_k-\varepsilon_{k^\prime}|\geq\hbar\omega$, these
terms are very small for $\hbar\omega\gg\delta\varepsilon$.
Therefore, they can be neglected if
$\hbar\omega\gg\delta\varepsilon$. Taking into account the
uncertainty relation for energy,
$\delta\varepsilon\sim\hbar/\tau$, this condition can be rewritten
in the form (\ref{tau}). Thus, both classical and quantum
description of the problem lead to the same result: The
collisional absorption (emission) of field energy by conduction
electrons can be neglected if the field frequency $\omega$ is high
enough.

Assuming the condition (\ref{tau}) to be satisfied, we can omit
terms with $n\neq0$ in eq.~(\ref{W}). Then the probability of
electron scattering (\ref{W}) takes the final form
\begin{equation}\label{W0}
w_{\mathbf{k}^\prime\mathbf{k}}=\frac{2\pi}{\hbar}J_0^2\left(f_{\mathbf{k}^\prime\mathbf{k}}\right)
\left|U_{\mathbf{k}^\prime\mathbf{k}}\right|^2
\delta(\varepsilon_{k^\prime}-\varepsilon_k).
\end{equation}
The formal difference between the scattering probability of
dressed electron (\ref{W0}) and the conventional expression for
the scattering probability of bare electron \cite{Landau_3},
\begin{equation}\label{L}
w^{(0)}_{\mathbf{k}^\prime\mathbf{k}}=\frac{2\pi}{\hbar}
\left|U_{\mathbf{k}^\prime\mathbf{k}}\right|^2
\delta(\varepsilon_{k^\prime}-\varepsilon_k),
\end{equation}
consists in the Bessel-function factor
$J_0^2\left(f_{\mathbf{k}^\prime\mathbf{k}}\right)$, where
$f_{\mathbf{k}^\prime\mathbf{k}}$ is proportional to the field
amplitude $E_0$ [see eq.(\ref{f})].  This factor arises from the
strong electron coupling to the field: If the field is absent
($E_0=0$), this factor is equal to unity and, as expected, the
expressions (\ref{W0}) and (\ref{L}) coincide. However, for
$E_0\neq0$ just this factor leads to the effect announced above.
Namely, it follows from the oscillatory behavior of the Bessel
function that there are amplitudes of the field, $E_0$, for which
the Bessel function $J_0(f_{\mathbf{k}^\prime\mathbf{k}})$ is
null. As a consequence, the field can turn the probability of
scattering (\ref{W0}) into zero. Particularly, eq.~(\ref{W0})
describes the probability of electron backscattering if
$\mathbf{k}^\prime=-\mathbf{k}$. This probability,
\begin{equation}\label{WW}
W=\frac{w_{-\mathbf{k}\mathbf{k}}}{w^{(0)}_{-\mathbf{k}\mathbf{k}}}=J^2_0\left(\frac{2e\mathbf{E}_0
\mathbf{k}}{m\omega^2}\right),
\end{equation}
is qualitatively pictured in fig.~2 as a function of the field
amplitude $E_0$. Evidently, the backscattering is absent for the
field amplitudes $E_0$ satisfying the equation
\begin{equation}\label{J}
J_0\left(\frac{2e\mathbf{E}_0\mathbf{k}}{m\omega^2}\right)=0
\end{equation}
which gives nulls of the probability (\ref{WW}).
\begin{figure}[th]
\begin{center}
\includegraphics[width=0.40\textwidth]{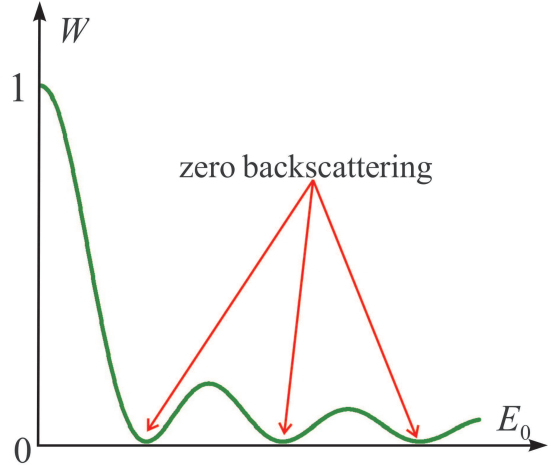}
\caption{The probability of backscattering of a dressed electron
as a function of amplitude of dressing field.}
\end{center}
\end{figure}

\section{Discussion and conclusions} It should be noted that
the expression (\ref{W0}) describes the probability of electron
scattering within the main (first) order of the perturbation
theory, which corresponds to the Born approximation
\cite{Landau_3}. As a consequence, the nulling of the probability
(\ref{W0}) suppresses the backscattering effectively if the
scattering potential $U(\mathbf{r})$ meets the well-known
conditions of applicability of this approximation \cite{Landau_3}.
These conditions cover a lot of physical situations which can be
experimentally realized in various conducting systems.
Particularly, the scattering of conduction electrons in modern
nanostructures arises from a very weak scattering potential
$U(\mathbf{r})$ and can be described by the theory elaborated
above. However, it can be useful for possible experiments to
estimate the remanent backscattering arisen from the second order
of the perturbation theory. Going in the same way as before, one
can derive the second-order probability of the electron
scattering,
\begin{eqnarray}\label{W2}
w_{\mathbf{k}^\prime\mathbf{k}}&=&\frac{2\pi}{\hbar}
\left|\sum_{\mathbf{k}^{\prime\prime}}J_0\left(f_{\mathbf{k}^{\prime}\mathbf{k}^{\prime\prime}}\right)J_0\left(f_{\mathbf{k}^{\prime\prime}\mathbf{k}}\right)\frac{U_{\mathbf{k}^\prime\mathbf{k}^{\prime\prime}}
U_{\mathbf{k}^{\prime\prime}\mathbf{k}}}{\varepsilon_k-\varepsilon_{k^{\prime\prime}}}\right|^2\nonumber\\
&\times&\delta(\varepsilon_{k^\prime}-\varepsilon_k),
\end{eqnarray}
which describes the remanent scattering in the high-frequency
limit if the main scattering (\ref{W0}) is nulled by a dressing
field. As expected, the probability (\ref{W2}) is negligibly small
under the conditions of the Born approximation \cite{Landau_3}.

Replacing the scattering probability of bare electron (\ref{L})
with the scattering probability of dressed electron (\ref{W0}) in
the conventional kinetic Boltzmann equation for conduction
electrons (see, e.g., ref.~\cite{Anselm}), one can calculate
stationary (dc) transport properties of dressed electrons.
Particularly, the discussed suppression of the electron
backscattering leads to increasing electron mobility. Applying a
stationary voltage to a conducting system dressed by a strong
high-frequency electromagnetic field, the field-induced decreasing
of electrical resistance can be measured experimentally. It should
be stressed that the probability (\ref{W0}) describes an elastic
scattering of dressed electron. However, in conducting systems
there are various mechanisms of electron scattering, including
both elastic scattering processes arisen from impurities and
unelastic ones caused by phonons. Therefore, the given theory can
accurately describe transport properties of dressed electrons if
the elastic scattering is dominant. This always takes place if
temperature is low enough.

The discussed effect is caused by nonlinear electron interaction
with a strong electromagnetic field. Mathematically, this follows
from the key expression (\ref{J}), where the field amplitude $E_0$
takes place in argument of highly nonlinear Bessel function.
Therefore, the effect under consideration cannot be derived from
the usual treatment of the problem, which is based on expansion of
the kinetic Boltzmann equation in a power series of field
amplitude. It should be noted also that the condition (\ref{tau})
requires a high field frequency $\omega$. On the other hand, if
the frequency $\omega$ is very high, the condition (\ref{J}) can
be realized only for very high field amplitudes $E_0$. Since high
field amplitudes correspond to high field intensities, strong
fields satisfying both the condition (\ref{tau}) and the condition
(\ref{J}) can fluidize a conductor. To avoid fusion of
experimental samples, one needs to use conductors with high
electron mobility (for instance, modern nanostructures), where the
conditions (\ref{tau}) and (\ref{J}) can be satisfied for low
field intensities.

Though the effect under consideration is of general character, it
will be most pronounced in one-dimensional nanostructures (for
instance, in semiconductor quantum wires, carbon nanotubes, etc)
since the backscattering is the only mechanism restricting the
mean free path of conduction electrons there. It should be noted
also that low-dimensional conductors are most preferable from
experimental viewpoint due to the absence of the screening of
dressing electromagnetic field. Therefore, let us estimate
observability of the discussed effect for conduction electrons in
one-dimensional conductors (for definiteness, in semiconductor
quantum wires with the effective electron mass $m\sim10^{-29}$~g).
For such conductors, the modern nanotechnologies allow to reach
the electron mobility $\mu=e\tau/m\sim10^7$~cm$^2/$V$\cdot$s that
corresponds to the electron relaxation time $\tau\sim10^{-9}$~s.
Therefore, the condition (\ref{tau}) can be satisfied for the
field frequencies $\omega\geq10^{10}$~Hz. Let the wire be filled
by a degenerate electron gas with the Fermi energy
$\varepsilon_F=\hbar^2k^2_F/2m\sim10^{-2}$~eV that corresponds to
the Fermi electron wave vector $k_F\sim10^6$~cm$^{-1}$. Assuming
the wire to be exposed to a high-frequency field (\ref{E})
directed along the wire, the condition (\ref{J}) can be satisfied
for the Fermi electrons if the field amplitude is
$E_0\geq10^{2}$~V/m. This field amplitude corresponds to the field
intensity $I\geq10^{-3}$~W/cm$^2$. Thus, the suppression of
electron backscattering can be realized for experimentally
reasonable parameters of the field. Finalizing the discussion,
another experimental consequence of this effect should be noted:
Since the factor
$J_0^2\left(f_{\mathbf{k}^\prime\mathbf{k}}\right)$ in the
scattering probability (\ref{W0}) is the oscillating function of
the field amplitude $E_0$ (see fig.~2), all stationary electronic
transport characteristics of field-dressed conductors are expected
to oscillate in the same manner.

Summarizing the aforesaid, the novel quantum effect
--- the nulling of the Born backscattering caused by a
high-frequency electromagnetic field
--- is claimed. It should be stressed that the conditions of the effect (\ref{tau}) and
(\ref{J}) depend only on field parameters and do not depend on
concrete form of the scattering potential. Thus, always there is
an electromagnetic field resulting in the nulling. As a
consequence, the effect is of universal physical nature. Since the
suppression of electron backscattering results in decreasing
electrical resistance, the solved quantum-mechanical problem opens
an unexplored way to control electronic transport properties of
conductors and electronic devices by a high-frequency
electromagnetic field. Particularly, the elaborated theory is
applicable to nanostructures exposed to a strong monochromatic
electromagnetic wave.

\section{Acknowledgements} The work was partially supported by the
Russian Ministry of Education and Science, RFBR (projects No.
13-02-90600 and No. 14-02-00033) and FP7 project QOCaN.

\end{document}